# Análisis de rendimiento y eficiencia energética en el clúster Raspberry Pi Cronos

Martha Semken[1], Mariano Vargas[1], Ignacio Tula[1], Giuliana Zorzoli[1], Andrés Rojas Paredes[1]

[1]Instituto de Ciencias (ICI) Universidad Nacional de General Sarmiento – UNGS

{msemken, avargas, itula, gjzorzoli, arojas}@campus.ungs.edu.ar

**Abstract.** En este artículo se presenta una evaluación del rendimiento computacional y la eficiencia energética del clúster Cronos, compuesto por microcomputadoras Raspberry Pi modelo 4 y modelo 3b orientado a fines educativos. Se realizaron pruebas experimentales utilizando el benchmark High Performance Linpack (HPL), bajo un entorno de gestión de recursos configurado con Slurm y comunicación paralela mediante Open MPI. El estudio se centró en analizar la escalabilidad, la estabilidad y el consumo eléctrico durante la ejecución de cargas intensivas de cómputo, considerando distintas configuraciones de nodos. Los resultados muestran que el clúster alcanza un rendimiento de hasta 6.91 GFLOPS en configuraciones homogéneas de 6 nodos RPi 4, y que el uso de nodos heterogéneos (incluyendo Raspberry Pi 3B) puede impactar negativamente en la estabilidad y la eficiencia. Asimismo, se midió el consumo eléctrico total del sistema durante las ejecuciones, permitiendo estimar la relación rendimiento/consumo (GFLOPS/W) como métrica comparativa. Este estudio constituye un aporte concreto para el diseño, evaluación y aprovechamiento de clústeres ARM de bajo costo en contextos educativos y de investigación.

1. **Introducción**

En el contexto actual de la computación, la demanda de soluciones eficientes y escalables para procesamiento paralelo ha impulsado la adopción de clústeres de bajo costo basados en arquitecturas ARM. En particular, las computadoras de placa única (Single Board Computer, SBC) como la Raspberry Pi ofrecen una alternativa accesible para explorar conceptos avanzados de paralelismo, distribución de tareas y eficiencia energética, tanto en ámbitos educativos como en proyectos de investigación aplicada (ver [1]).

El clúster Cronos, desarrollado en el Instituto de Ciencias de la Universidad Nacional de General Sarmiento (UNGS), está compuesto por microcomputadoras Raspberry Pi 3b y 4; fue concebido con fines didácticos, orientado a la enseñanza de cómputo paralelo. Este trabajo se inscribe en el marco de un proyecto de investigación en curso, que tiene como objetivo la evaluación de un clúster de computadoras utilizando computadoras Raspberry Pi, evaluar el rendimiento, la escalabilidad y la eficiencia energética de sistemas de cómputo de bajo costo, con potencial para ser utilizados en entornos académicos y experimentales.

En etapas previas se abordó el diseño físico del clúster, su ensamblaje, y la instalación del software base, incluyendo herramientas como Munge (autenticación ver [2]), Slurm (gestión de recursos, ver [3])) y Open MPI (comunicación paralela). Sobre esta infraestructura funcional, se realizó una evaluación experimental utilizando el benchmark High Performance Linpack (HPL; Dongarra, Luszczek, & Petitet, 2003 ver [4]), para medir el rendimiento explorando distintas configuraciones de nodos y parámetros computacionales en términos de FLOPS (Floating Point Operations Per Second), una unidad que expresa la cantidad de operaciones en punto flotante que un sistema puede realizar por segundo.

Asimismo, se llevaron a cabo mediciones de corriente y tensión durante la ejecución de los trabajos, lo que permitió estimar el consumo eléctrico total y calcular métricas de eficiencia como GFLOPS/Watt utilizadas en rankings como el Green500 (2024). Esto se enmarca en los principios del Green Computing, que promueven el desarrollo de soluciones computacionales de bajo impacto ambiental (ver [5]).

El objetivo principal de este trabajo es analizar la relación entre rendimiento y consumo energético en un entorno heterogéneo, accesible y replicable, y reflexionar sobre su viabilidad como plataforma de experimentación en contextos educativos y de investigación. A su vez, se documentan los principales problemas técnicos encontrados durante las pruebas, junto con las soluciones implementadas, con el fin de aportar conocimiento práctico a la comunidad interesada en el desarrollo de clústeres ARM de bajo consumo.

## 2. Líneas de investigación y desarrollo

El presente trabajo se enmarca en un proyecto que se estructura en tres fases principales, cada una con objetivos, procedimientos y productos específicos. Esta división permite abordar de forma progresiva el diseño, implementación, evaluación y aprovechamiento del clúster Cronos.

- **Primera Fase**: Configuración inicial y pruebas funcionales
  Objetivo: montaje del hardware, instalación del sistema operativo, configuración de red, NFS, Munge, Slurm y Open MPI.
  Estado: Completada. Se documentó en el artículo "Configuración y Puesta en Marcha del Clúster Raspberry Pi Cronos" (CACIC 2024), donde se verificó el funcionamiento básico del clúster y su capacidad para ejecutar tareas paralelas simples (ver [6]).
  Actividades destacadas: en diciembre de 2024 se llevó a cabo un workshop de introducción al cómputo paralelo utilizando el clúster Cronos. En dicho taller, más de cuarenta estudiantes de la carrera de Programación adquirieron conocimientos sobre el diseño de algoritmos paralelos y realizaron prácticas en lenguaje C utilizando herramientas y metodologías propias del cómputo paralelo. Durante las actividades se identificaron limitaciones relacionadas con la administración de usuarios, la escalabilidad y el rendimiento del almacenamiento, lo que motivó la implementación de mejoras en la infraestructura del clúster.

- **Segunda Fase**: Evaluación del rendimiento y eficiencia energética
  Objetivo: medir el rendimiento computacional del clúster utilizando benchmarks estandarizados (HPL), analizar su escalabilidad con diferentes configuraciones de nodos y evaluar el consumo energético real durante la ejecución de tareas intensivas.
  Estado: En desarrollo. Este artículo documenta los avances de esta fase, presentando resultados detallados de pruebas con diferentes parámetros, nodos y condiciones de operación.
- **Tercera Fase**: Desarrollo de aplicaciones paralelas y documentación técnica.
  Objetivo: diseñar e implementar aplicaciones que aprovechen el entorno del clúster, generar material didáctico, guías de replicación y optimización del sistema para futuros usos educativos o investigativos.
  Estado: a planificarse tras la finalización de las pruebas de rendimiento.
  Perspectivas: se prevé continuar desarrollando cursos, prácticas y talleres de cómputo paralelo utilizando Cronos como plataforma formativa. Estas actividades permitirán seguir testeando nuevas aplicaciones, incorporar mejoras al entorno y fortalecer la enseñanza de programación paralela en la carrera de Informática.

La presente publicación se focaliza en los resultados obtenidos durante la Fase II, centrada en la evaluación del rendimiento computacional y el análisis del consumo energético. A lo largo de esta etapa se realizaron diversas pruebas utilizando la herramienta High Performance Linpack (HPL), junto con mediciones de corriente eléctrica para estimar la eficiencia del sistema en términos de GFLOPS/Watt. También se analizó el impacto de diferentes configuraciones de nodos (homogéneas vs. heterogéneas) y se documentaron incidentes técnicos surgidos durante la experimentación, junto con las estrategias aplicadas para su resolución.

## 3. Evaluación del rendimiento y eficiencia energética

Esta sección presenta los resultados experimentales obtenidos al ejecutar el benchmark High Performance Linpack (HPL) en el clúster Cronos, con el objetivo de medir su rendimiento computacional (en GFLOPS) y evaluar su comportamiento bajo distintas configuraciones. Las pruebas se realizaron utilizando herramientas estándar del entorno científico, con un enfoque en la reproducibilidad y la estabilidad de las ejecuciones.

### 3.1 Introducción al benchmark HPL Linpack

High Performance Linpack (HPL) es una implementación paralela y portátil del benchmark Linpack, utilizado internacionalmente para evaluar la capacidad de cómputo en punto flotante de alto rendimiento en clústeres y supercomputadoras. Linpack resuelve sistemas de ecuaciones lineales densos del tipo Ax=b, donde A es una matriz cuadrada de coeficientes, x y b son vectores. Esta operación es fundamental en muchas aplicaciones científicas, y su resolución eficiente permite medir el desempeño real de una arquitectura computacional.

HPL realiza esta tarea utilizando aritmética de doble precisión (64 bits), aprovechando bibliotecas como BLAS (ver [7]) para operaciones numéricas intensivas, y MPI para la comunicación entre múltiples nodos en sistemas distribuidos. El benchmark permite ajustar diversos parámetros de ejecución, entre ellos:

- N: tamaño del sistema (dimensión de la matriz A)
- NB: tamaño de bloque para la factorización

P x Q: distribución de procesos en una cuadrícula bidimensional, donde P y Q definen cómo se reparten los procesos MPI entre las filas y columnas lógicas de la topología. Esta distribución PxQ permite adaptar el mapeo de procesos a la arquitectura del clúster, maximizando la eficiencia de las comunicaciones y el uso de la memoria.

El resultado más relevante del benchmark es el rendimiento expresado en GFLOPS (giga operaciones en punto flotante por segundo), que representa la tasa de cómputo alcanzada durante la resolución del sistema.

HPL se utiliza como métrica estandarizada de rendimiento en computación de alto desempeño, y es la herramienta empleada por el ranking internacional TOP500 para evaluar y comparar las supercomputadoras más poderosas del mundo (ver [8]).

En este trabajo, HPL se utilizó como herramienta principal para caracterizar el comportamiento del clúster Cronos bajo cargas intensivas. Los resultados obtenidos fueron luego relacionados con el consumo eléctrico del sistema, permitiendo así estimar su eficiencia energética (GFLOPS/Watt). Este enfoque se alinea con las tendencias actuales del Green Computing y la evaluación sostenible de infraestructuras informáticas.

### 3.2 Resultados destacados con 6 nodos

Las pruebas más estables se lograron utilizando únicamente los 6 nodos Raspberry Pi 4 (4 GB de RAM), excluyendo a los Raspberry Pi 3B. Esta configuración homogénea permitió un mejor aprovechamiento de recursos, mayor estabilidad operativa y menor dispersión en los resultados.

| Nodos | N | NB | PxQ | Tiempo(s) | GFLOPS | Comentario |
|---|---|---|---|---|---|---|
| 6 | 3000 | 224 | 3x2 | 2872.55 | **6.2667** | Mejor resultado registrado |
| 6 | 3000 | 224 | 3x2 | 2889.50 | 6.2299 | Ejecución estable con fuentes de PC |
| 6 | 3000 | 224 | 3x2 | 2946.40 | 6.1096 | Primeras pruebas estables con fuente switch |

Se verificó que la combinación P=3, Q=2 es óptima para esta topología de 6 nodos, y que el tamaño de bloque NB = 224 maximiza el rendimiento sin comprometer la estabilidad. La mejor ejecución alcanzó **6.2667 GFLOPS**, estableciendo un techo de rendimiento para este entorno homogéneo.

Para esta ejecución, se midió la corriente consumida durante 35 minutos en intervalos de 5 minutos, utilizando una pinza amperimétrica conectada a la fuente switching de 5V. Aplicando el método del trapecio sobre los valores registrados, se estimó un consumo total de **0,448 Wh**. Considerando un rendimiento de **6.2667 Gflops**, la eficiencia energética calculada para esta prueba fue de aproximadamente **15,42 Gflops/Watt**. Los datos crudos y el gráfico correspondiente están disponibles en el repositorio complementario del proyecto (ver [12]).

Se volvieron a repetir los experimentos donde el script de Slurm se configuró con una tarea por nodo, por lo que en seis nodos se ejecutaron seis procesos en total. Esta configuración se repitió tres veces, obteniéndose un rendimiento medio de **6.19 Gflops** con una desviación muestral de ±0.07, mientras que el tiempo de ejecución promedió 2910 segundos con una desviación muestral de ±30.3. La desviación muestral se calculó como la raíz cuadrada de la varianza muestral, es decir, la suma de los cuadrados de las diferencias respecto de la media dividida por n−1, donde n es la cantidad de repeticiones. Esta medida cuantifica la dispersión de los valores obtenidos y constituye una estimación de la desviación estándar poblacional, recomendada en experimentos con un número reducido de corridas.

Posteriormente, dado que cada Raspberry Pi 4 dispone de cuatro núcleos de procesamiento, se modificó el script incorporando el parámetro **--ntasks-per-node=4**, lo que permitió aprovechar el paralelismo intra-nodo. Con esta nueva configuración, que asigna cuatro procesos por nodo (24 procesos en total), también se efectuaron tres repeticiones. El rendimiento mejoró significativamente: el valor medio alcanzado fue de **14.48 Gflops** con una desviación muestral de ±0.26, reduciendo el tiempo de ejecución a 1243 segundos con una desviación muestral de ±22.5. Este resultado representa un incremento superior al 130% respecto de la configuración inicial y constituye el mejor rendimiento medido (achieved performance) del clúster en las condiciones de prueba.

### 3.3 Resultados con 8 nodos y comparación

También se realizaron pruebas con 8 nodos (6 RPi 4 + 2 RPi 3B). Aunque podría esperarse un aumento de rendimiento, los resultados mostraron lo contrario: la diferencia fue mínima o incluso negativa, dependiendo de la configuración.

| Nodos | N | NB | P×Q | Tiempo (s) | GFLOPS | Comentario |
|---|---|---|---|---|---|---|
| 8 | 30000 | 160 | 3×2 | 2932.81 | **6.1379** | Igual distribución P×Q que con 6 nodos |
| 8 | 30000 | 256 | 8×1 | 6873.31 | 2.619 | Configuración forzada, rendimiento pobre |

Aunque en uno de los casos con 8 nodos se logró 6.1379 Gflops, el valor es solo 0.4% superior al obtenido con 6 nodos. Además, las pruebas con 8 nodos mostraron menor estabilidad y varios incidentes técnicos que comprometen su uso prolongado bajo carga.

Durante las ejecuciones de HPL se presentaron distintos incidentes técnicos que afectaron la estabilidad del clúster. Por ejemplo, se detectaron fallas de comunicación entre nodos debido a la desincronización horaria provocada por la falta de acceso a servidores NTP, lo cual impidió la autenticación correcta con SLURM (mensaje "Zero Bytes were transmitted or received"). En otro caso, el nodo7 (RPi 3B) dejó de aceptar tareas pese a responder a ping, lo que interrumpió temporalmente la ejecución del job. La intervención manual con scontrol y el reinicio físico del nodo permitieron continuar la ejecución sin pérdida de datos. Estos episodios motivaron decisiones técnicas clave, como la exclusión de los nodos RPi 3B en las pruebas finales, privilegiando configuraciones más estables y homogéneas. Se puede ver en la figura 1 los resultados finales de las pruebas.

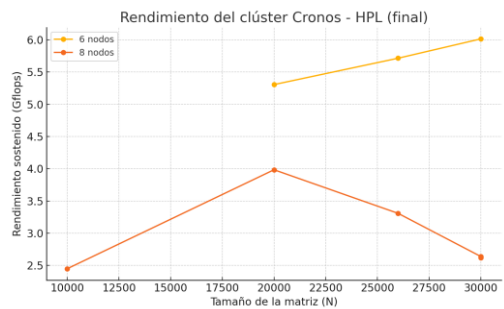

Figura 1: Resultados finales de HPL

### 3.4 Escalabilidad

En términos de escalabilidad, el clúster mostró un comportamiento positivo hasta los 6 nodos homogéneos Raspberry Pi 4, con un aumento sostenido del rendimiento al incrementarse el número de procesos. Sin embargo, al extender la ejecución a los 8 nodos disponibles (6 RPi 4 y 2 RPi 3), los experimentos con HPL en problemas de gran tamaño se interrumpieron repetidamente debido a fallas de estabilidad en los nodos RPi 3. Esta situación impidió obtener resultados comparables y pone en evidencia las limitaciones que introduce la heterogeneidad del hardware.

De este modo, puede afirmarse que el clúster es escalable hasta 6 nodos homogéneos, mientras que la inclusión de nodos menos potentes genera un cuello de botella que compromete tanto el rendimiento como la estabilidad. En futuras etapas, cuando se complete la migración a un clúster homogéneo de RPi 4, será posible repetir las pruebas y caracterizar con mayor precisión la relación entre el número de nodos, el rendimiento medido en Gflops y la eficiencia energética.

### 3.5 Comparación con estudios similares

Los resultados de Cronos pueden situarse en el marco de investigaciones previas sobre clústeres basados en hardware de bajo costo y computadoras SBC. En particular, Abarca Calderín (2020) analizó un clúster heterogéneo compuesto por Raspberry Pi 3 y Raspberry Pi 4, encontrando que las configuraciones homogéneas no solo simplifican la operación, sino que también ofrecen un rendimiento y estabilidad superiores, lo cual coincide con nuestra decisión de prescindir de los nodos Raspberry Pi 3 (ver [9]). Por su parte, Rodríguez Eguren, Chichizola y Rucci (2018) reportaron que, si bien este tipo de clústeres permite demostrar conceptos de cómputo paralelo y eficiencia energética, los valores absolutos de rendimiento están acotados por las limitaciones del hardware SBC (ver [10]). Finalmente, Cruz Salazar (2021) exploró un clúster de Raspberry Pi orientado principalmente a fines educativos, destacando la facilidad de implementación y el bajo costo, pero con cargas de trabajo más livianas (ver [11]).

En este contexto, Cronos —con seis nodos homogéneos Raspberry Pi 4 y una configuración optimizada de cuatro tareas por nodo— alcanza un rendimiento promedio de 14,48 GFLOPS (±0,26), superando claramente los resultados obtenidos en configuraciones heterogéneas y posicionándose como un aporte relevante en términos de costo/rendimiento dentro de esta línea de investigación.

### 3.6 Consistencia en el uso de recursos, con balance de carga uniforme.

El uso exclusivo de nodos homogéneos Raspberry Pi 4 permitió asegurar un balance de carga uniforme y una menor complejidad operativa, al evitar las adaptaciones requeridas por las diferencias de hardware. En particular, los RPi 3B introducían desincronizaciones y posibles cuellos de botella de red o memoria, lo que comprometía la estabilidad del sistema en ejecuciones de gran escala.

Esta decisión técnica no solo permitió obtener resultados más confiables y representativos del potencial real del clúster en condiciones controladas, sino que además estableció un punto de partida sólido para futuros análisis de escalabilidad en configuraciones homogéneas, donde será posible evaluar con mayor precisión la relación entre número de nodos, rendimiento y eficiencia energética.

## 4. Aplicaciones paralelas y validación funcional del clúster

Se realizaron pruebas aplicando MPI y MPI + OpenMp sobre un algoritmo para el cálculo de PI, aplicando la regla del punto medio. En base a las mediciones realizadas se obtuvieron los siguientes gráficos.

Con MPI

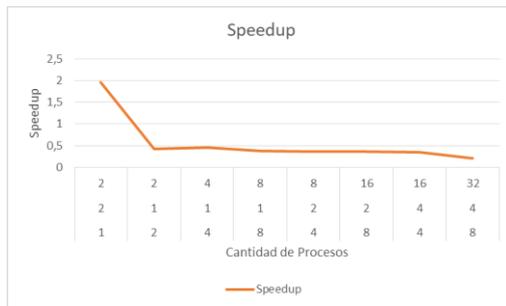 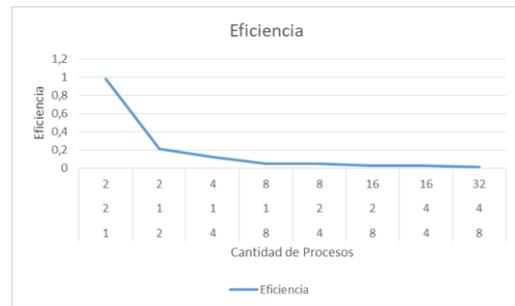

Figura 2: Speedup MPI          Figura 3: Eficiencia MPI

Con MPI + OpenMP

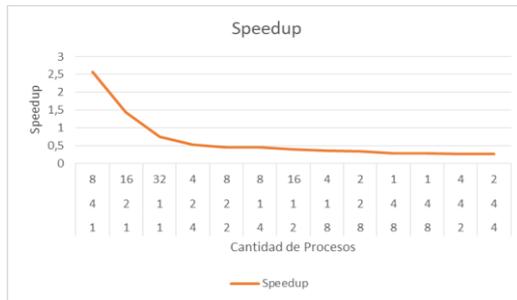 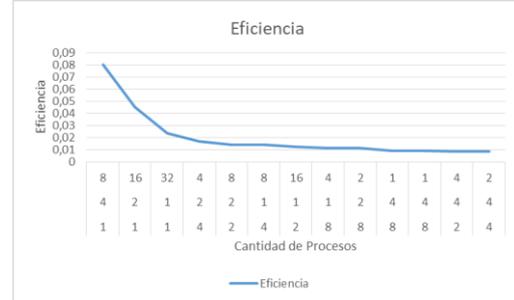

Figura 4: Speedup MPI          Figura 5: Eficiencia OpenMP + MPI

En MPI se ve mayor disminución de eficiencia al aumentar el número de procesos, las mediciones con MPI+OpenMP (modelo híbrido) se realizaron siempre con 32 procesos pero con distintas configuraciones de nodos, procesos e hilos. El modelo híbrido escala mejor, si se aprovechan todos los cores por nodo, no teniendo la misma eficiencia cuando se usan muchos nodos y solo un core por cada uno.

Todos los scripts, archivos de configuración (como HPL.dat), códigos fuente en C utilizados para el cálculo del número π en versiones secuencial, MPI e híbrida, así como los resultados obtenidos (tiempos, GFLOPS, consumo eléctrico, cálculos de speedup y eficiencia), se encuentran disponibles en el repositorio público GitHub del proyecto de investigación (ver [12]).

## 5. Conclusiones

Los experimentos realizados muestran que, con un ajuste adecuado de configuración, un clúster de bajo costo basado en Raspberry Pi puede duplicar su rendimiento efectivo y sostener ejecuciones estables bajo carga. Si bien en la discusión inicial se compararon nuestros valores con los reportados en iniciativas como Green500, reconocemos que este contraste tiene un carácter meramente ilustrativo, ya que involucra tecnologías muy diferentes.

En términos de relevancia educativa y de investigación aplicada, resulta más apropiado comparar Cronos con clústeres de PCs comunes utilizados en laboratorios universitarios. Estudios como los de Cruz Salazar (2021) y Rodríguez Eguren et al. (2018) evidencian que dichos entornos ofrecen un mayor rendimiento absoluto por nodo, pero a un costo energético y económico más elevado. Así, Cronos se ubica en un punto intermedio: su rendimiento no compite con un clúster de PCs, pero su bajo consumo, costo reducido y facilidad de replicación lo convierten en una plataforma valiosa para la enseñanza y experimentación.

Como trabajo futuro inmediato, se plantea la realización de una comparación sistemática con un clúster de PCs de laboratorio, con el fin de evaluar de mas forma más justas la relación rendimiento/consumo/costo entre ambas plataformas.